# Metasurfaces for Near-Eye Augmented Reality

*Shoufeng Lan,[†,‡] Xueyue Zhang,[†,‡] Mohammad Taghinejad,[†] Sean Rodrigues,[†,§] Kyu-Tae Lee,[†] Zhaocheng Liu,[†] and Wenshan Cai[†,§,]\**

[†] School of Electrical and Computer Engineering, Georgia Institute of Technology, Atlanta, Georgia 30332

[§] School of Materials Science and Engineering, Georgia Institute of Technology, Atlanta, Georgia 30332

[‡] Authors contributed equally to this work

\* Correspondence should be addressed to W.C. (wcai@gatech.edu)

ABSTRACT: Augmented reality (AR) has the potential to revolutionize the way in which information is presented by overlaying virtual information onto a person's direct view of their real-time surroundings. By placing the display on the surface of the eye, a contact lens display (CLD) provides a versatile solution for compact AR. However, an unaided human eye cannot visualize patterns on the CLD simply because of the limited accommodation of the eye. Here, we introduce a holographic display technology that casts virtual information directly to the retina so that the eye sees it while maintaining the visualization of the real-world intact. The key to our design is to introduce metasurfaces to create a phase distribution that projects virtual information in a pixel-by-pixel manner. Unlike conventional holographic techniques, our metasurface-based technique is able to display arbitrary patterns using a single passive hologram. With a small form-factor, the designed metasurface empowers near-eye AR excluding the need of extra optical elements, such as a spatial light modulator, for dynamic image control.

Metasurfaces enable arbitrary wavefront shaping with unprecedented flexibility by producing controllable abrupt changes in the phase, amplitude, and polarization of light waves at a thickness comparable or smaller than the wavelength of light.[1-4] To enable such spatial variations in the optical responses, metasurfaces employ geometric phase,[5,6] coupled modes,[7,8] or dipolar resonances of miniature light scatterers.[9-12] With their algorithmic arrangement of two-dimensional (2D) scatterers, metasurfaces have demonstrated a unique ability to manipulate light for a wide range of applications including wave concentration, polarization control, and beam steering, all with an ultra-compact device footprint within optically thin films.[13-18] Among them, metasurface-based holography is a promising display technique which provides a high resolution and low noise image that corresponds to a static hologram by precisely reconstructing the amplitude and phase information of light with high efficiency.[19-22] Because most optical imaging components are essentially shaping the wavefront of light to achieve a desired function, metasurfaces can replace the role of these conventional lenses and mirrors for image reconstruction,[23-25] thus, making possible the reduction of the dimension of image systems such as augmented reality (AR).

AR is revolutionizing the way information is presented to people by overlaying virtual information such as graphics and captions onto a direct view of a person's real-world surroundings.[26-28] A central component of AR systems is a display that combines real and virtual images so that both are seen at the same time.[29] Among the ideation of current display technologies for AR, optical see-through displays stand out since they leave the real-world perception largely intact and display the AR overlay by means of beamsplitters and miniature projection lenses.[30] The current prevailing technology for this purpose is in the form of head-mounted displays (HMDs), which are bulky and offer a limited field of view (FOV). Furthermore, HMDs substantially hinder the social acceptance of AR in everyday life owing to the appearance of the user and the privacy concern of others. One approach to remove these hurdles is bringing the display closer to the eye and integrating it with a contact lens.[31]

A contact lens display (CLD) is a promising candidate of the "ultimate display" for AR, per Ivan Sutherland's vision in 1965,[32] because it is imperceptible to those around the user. By virtue of circuitry, telecommunication, nanotechnologies, and microelectromechanical systems (MEMS), a display system integrated in a contact lens is feasible.[33] In recent years, a single pixel display using a wireless-powered light-emitting diode (LED) was first developed on a contact lens.[34] The display was then tested in the eye of a rabbit showing no adverse effects.[35] Later, a spherically deformed liquid crystal cell was integrated into a contact lens contributing more to a realistic CLD.[36] Before envisioning the fascinating CLD-based AR, however, we must solve a fundamental optical problem that the displayed patterns on the corneal surface cannot be faithfully observed by the retina because of the finite accommodation of the human eye. Initial attempts to solve this problem integrate a micro-Fresnel lens to focus light from a single pixel,[35] however, displaying 2D patterns is still challenging.

In this work, we tackle this challenge by employing a metasurface to holographically cast virtual information, from 2D patterns on a CLD, onto the fovea region of the retina in a pixel-by-pixel manner. This technique holographically casts information that locates on the surface is in distinct contrast to that of a meta-lens because of the zero-distance between the object and the lens excludes the formulating of images governed by the well-known lens maker's formula. With proper accommodation, the human eye can perceive the pixelated patterns appearing at the near point. The metasurface, which generates the predesigned phase distribution using silicon nanobeams, occupies only 1% of the pupil area, leaving the visualization of the real-world environment intact. By superimposing the virtual and unaffected real-world information, the metasurface-based display technique allows the realization of AR on a contact lens. The metasurface features the smallest form factor, adding a sub-micrometer thickness and a sub-microgram weight to a normal contact lens.

The main objective behind AR is to superimpose a virtual graphic or caption onto the real-world environment such that the two visuals are detected by the retina simultaneously, thereby augmenting the presence of the real-world with external information. At the intersection of this virtual and real-world is the display. Here, we introduce a metasurface-based display technique, as shown in Figure 1a, with a miniature footprint for AR on a contact lens. Because of the limited accommodation of the eye, patterns embossed on the surface of the contact lens, which sits at the cornea, cannot be seen on the retina by an unaided human eye. We resolve this problem by introducing a metasurface to holographically cast the patterns onto the fovea region of the eye. With proper accommodation, the retina sees the displayed virtual information and the real-world environment simultaneously. Figure 1b shows that the virtual information will be formed by a coherent light and a metasurface situated immediately above the virtual information. The metasurface produces a phase distribution that holographically generates an enlarged image of the pattern. When fabricated on a contact lens, the metasurface chip will occupy less than 1% of the pupil area, and therefore will have a negligible impact on the normal operation of the eye. As such, normal functions of the human eye, including its natural field of view, perception of color, stereoscopic ability, real-time responses, and so on, are perfectly preserved.

The metasurface in our design is a dielectric gradient metasurface composed of dense arrangements of silicon nanobeams (Figure 1b, right). These nanobeams serve as dielectric optical antennas that support a series of Mie resonances under illumination of polarized light. The scattered waves of the two input polarizations, with an electric field parallel and perpendicular to the nanobeams, exhibit substantial phase retardation, and thusly lead to a strong and controllable birefringent behavior. When properly designed, a nanobeam can act as a local half waveplate with a relative phase retardance of π between the two primary polarizations. As a result, such silicon nanobeams serve as Pancharatnam-Berry (PB) phase elements with geometrically induced abrupt phase change.[5,6] The magnitude of the induced PB-phase covers from −π to π, and its value is determined by the specific orientation of the silicon nanobeams. By manipulating the PB-phase of light from a display, the metasurface manages to holographically project the patterns towards the eye. To control the propagation of light in an eye by using metasurfaces, however, we must know the accurate optical properties of all parts of the eye as well as the accommodation associated with the muscle movement of the individuals, which is highly challenging, if not impossible. Fortunately, the metasurface hologram is designed to have a conjugate image against the eye, which is in the air with well-defined optical properties. Thus, we can design the patterns on the retina by aiming the phase distribution on the metasurface for a targeted conjugate image in the air, so that the eye sees the displayed virtual information as if it were at the location of the conjugate image, thanks to the intelligent accommodation ability of the human eye.

Unlike conventional holographic techniques where a phase distribution from a hologram only corresponds to a fixed image, our metasurface generates a phase distribution that forms enlarged images corresponding to arbitrary patterns on the display. Figure 2a shows that the metasurface with a size of 500 μm by 500 μm, situated at the center of the contact lens, projects virtual information, the letter "F," onto the fovea region (5 mm by 5 mm) of the retina. With the help of the accommodation, the eye perceives the letter "F" as if it is the enlarged and pixelated conjugate image, 5 cm by 5 cm in size, 25 cm away from the surface of the contact lens (at the near point). In this way, the virtual information displayed on the surface of the cornea can be superimposed onto the image of the real-world environment on the retina. This would not be possible, without the phase manipulating ability of the metasurface.

To holographically transform the pixelated patterns, we design the phase-retardance of the metasurface in the spatial frequency domain, where linear algebra can be performed on the phase delay from all optical components. By applying the Gerchberg-Saxton algorithm,[37] in which Fourier and inverse Fourier transforms iterate between the two optical planes involved, we obtain the phase profile for an array of 10 by 10 pixels in Figure 2b. Each pixel, 50 μm by 50 μm in size, needs to possess a prescribed phase distribution, which is determined by both the system configuration and the specific coordinate location of the pixel. We discretize the PB-phase from $-\pi$ to $\pi$ into eight values with a step size of $\pi/4$ so that it can be experimentally realized while not compromising much of the conversion efficiency. We further verify the performance of the device with the discretized phase profile by numerically calculating the intensity distribution on the retina corresponding to that displayed on the contact lens. Figure 2c shows the 10 by 10 pixels that will be projected into the retina. As shown in Figure 2d, the retinal detector visualizes the virtual information, the letter "F" in this case (inset), generated by the display with the same phase profile as that for Figure 2c on the contact lens. In contrast, under the same illumination conditions, the retinal detector can only visualize a focused beam spot without the predesigned phase distribution and a randomized intensity distribution with a random phase profile (Figure S1). These results verify the validity of the phase distribution for holographically forming an image that is viewed on the retina in a pixel-by-pixel manner.

The phase distribution of the metasurface relies on the modification of the space-variant geometric phase, or the PB-phase, that associates with the polarization rather than the conventional phase that results from the path difference of light. We express the polarization of the output light from the nanobeams in a helicity basis as the following.

$$|E_{out}\rangle = \sqrt{\eta_E}\,|E_{in}\rangle + \sqrt{\eta_R}\,e^{i2\theta(x,y)}|E_R\rangle + \sqrt{\eta_L}\,e^{-i2\theta(x,y)}|E_L\rangle$$

where $E_{out}$, $E_{in}$, $E_R$, and $E_L$ are the electric field of the output, input, right- and left-circularly polarized light, respectively; $\theta(x, y)$ is the space-variant angle of the nanobeams; and $\eta_E$, $\eta_R$, and $\eta_L$ are the transmission coefficients with $\eta_E = \left|\frac{1}{2}(t_{TE} + t_{TM}e^{i\Delta\phi})\right|^2$, $\eta_R = \left|\frac{1}{2}(t_{TE} - t_{TM}e^{i\Delta\phi})\langle L|E_{in}\rangle\right|^2$, and $\eta_L = \left|\frac{1}{2}(t_{TE} - t_{TM}e^{i\Delta\phi})\langle R|E_{in}\rangle\right|^2$, in which $\Delta\phi$ is the phase difference between the two polarizations (TE and TM) that are parallel and perpendicular to the nanobeams. By properly designing the geometry of the nanobeams, $t_{TE} = t_{TM}$ and $\Delta\phi = \pi$, the input light is fully converted into a cross circular polarization with a phase twice as large as the angle of the nanobeams.

We experimentally realize the phase distribution by designing the geometry of the silicon nanobeams at the specific location. First, we deposit a layer of 155 nm thick polysilicon on a fused silica substrate. Using this surface, we extract the refractive index of the polysilicon film by performing ellipsometry measurements (Figure S2). We then import the optical properties into a simulation package to determine the geometric parameters, including the periodicity and width, of the nanobeams by numerically simulating the transmission ($t_{TE}$ and $t_{TM}$) and phase ($\Delta\phi$) of light that passes through the nanobeams (Figure S3a and S3b). The resultant silicon nanobeams must faithfully produce circularly polarized light, such that we accomplish the condition where $t_{TE} = t_{TM}$ and $\Delta\phi = \pi$ at the operating wavelength of 543 nm. We optically characterize the fabricated nanobeams, with the predesigned geometric parameters (the periodicity is 230 nm and the width of the nanobeam is 70 nm), using an interferometry technique described in our previous paper.[38]

Figure S3 shows that the measured transmission (Figure S3c) and phase (Figure S3d) match the simulations (Figure S3a and S3b) relatively well; the mismatch between the predicted and experimental responses are due to fabrication and characterization uncertainties. Nevertheless, the characterizations of transmission and phase for TE and TM polarizations provide a guide for optimizing the conversion efficiency with metasurfaces using PB phases.

With the predesigned phase distribution generated from the silicon nanobeams, we demonstrate a metasurface-based display technique for applications of near-eye AR. Figure 3a is an SEM image of the fabricated metasurface. The silicon nanobeams are oriented at specific angles to realize the phase distribution required to holographically cast virtual information. During the experiment, we use an image of a photomask, instead of a real see-through display, to generate the virtual information. The virtual information and the real-world environment are viewed simultaneously by a retina camera with an eye-lens in front as shown in Figure S4. When all one-hundred pixels emit light (543 nm) forming the green uniform square in Figure 3b underneath the metasurface, the retina camera sees the whole information in a pixel-by-pixel manner (Figure 3c), experimentally confirming the validity of the metasurface-based display technique. To further testify the capability of our metasurface hologram to display arbitrary patterns, we generate multiple bar-patterns (Figure 3d - 3g) by moving the image of the photomask from one side of the metasurface to the other. The retina camera recognizes the moving bars in Figure 3h to 3k corresponding to the displayed patterns in Figure 3d to 3g, respectively.

More importantly, when the display generates a virtual graphic by emitting light at specific pixels, for example the pixels for the letter "F" (inset) in this case, the retina camera observes the enlarged, pixelated pattern in Figure 4a. The bright cross at the center is the zeroth-order diffraction, commonly seen in metasurface devices using geometric phases, owing to the imperfectly fulfilled condition, $t_{\text{TE}} = t_{\text{TM}}$ and $\Delta\phi = \pi$, together with the discretization of the phase distribution. The zeroth-order diffraction can be experimentally eliminated by placing an analyzer that is crossly polarized to the input light. The metasurface, with square side lengths of 500 μm by 500 μm, only occupies ~1% of the area of the pupil. This leaves the overall functionality of the eye unaffected and enables overlaying of the virtual and real-world information at the retina. Indeed, the retina camera precisely records the virtual information (Figure 4a) and the real-world environment (Figure 4b) simultaneously in Figure 4c, experimentally demonstrating the metasurface-based display technique towards AR on a contact lens.

These results clearly confirm the validity of the proposed method: a metasurface can holographically cast virtual information generated by a see-through display onto the retina, making possible AR on a contact lens. While the current design demonstrates relatively low resolution, augmenting the physical environment with simple annotation text and icons is known to be valuable for various applications such as warning indicators and visual cues in a navigation system. Furthermore, by designing more pixels for the display, for example 100 by 100 pixels, we verify via numerical simulation that our technique displays complex virtual information, such as a map of America (Figure S5), on a contact lens. We also note that stereoscopic binocular displays for better depth perception are also possible with the proposed system, as the metasurface chip can be individually engineered for each eye.

In summary, we experimentally demonstrated AR on a contact lens with a metasurface that enables the direct visualization of virtual information in a see-through display on the surface of the human eye. Our method of holographically casting the virtual information in a pixel-by-pixel manner provides a versatile solution to significantly minimize the footprint of AR devices. This feature makes AR imperceptible to neighboring people, which could make the CLD-based AR more socially accepted by the public. Moreover, the passive ultrathin metasurface allows for integration with current techniques and systems, such as a wearable display, for projecting images from a two-dimensional plane to free space without extra optical components, such as a spatial light modulator, for dynamic image control.

**Methods**

**Numerical simulation.** The transmission and phase retardance of silicon nanobeams were simulated using a commercial FEM solver (COMSOL). The optical properties of the polysilicon film for the simulation were obtained from ellipsometry measurements, using a Woollam ellipsometer. Linearly polarized light was applied as the input to the simulation area with periodic boundary conditions on the left and right and perfectly match layers on the top and bottom.

**Fabrication.** The polysilicon film was deposited on a fused silica substrate using the Tystar Poly Furnace at a temperature of 610 °C and a pressure of 25 mTorr. A thin layer of HSQ was spun on top of the polysilicon film to be used as the e-beam resist for e-beam lithography. After e-beam lithography pattering, reactive ion etching (RIE) was performed with an STS ICP standard oxide etcher for transferring the patterns to the polysilicon film. The polysilicon on the back of the substrate was removed by a Xactix Xenon difluoride etcher, while the front was protected by a photoresist.

**Optical characterization.** The transmission of the nanobeams was measured with a homemade setup dedicated to the spectral measurement of samples with microscopic dimensions. A broadband light source (B&W Tek BPS120) was used to illuminate the sample. A set of linear polarizers and waveplates was employed to control the polarization state of the incident light. The transmitted signal from the sample was collected by an objective lens (Mitutoyo, 100 × Plan Apo

NIR infinity-corrected) and transformed by another lens to form a magnified image of the sample. The transmission was obtained by normalizing the transmitted light from the desire area to that of the substrate. The phase delay between the TE and TM polarizations was extracted with a polarization interferometric technique.

...

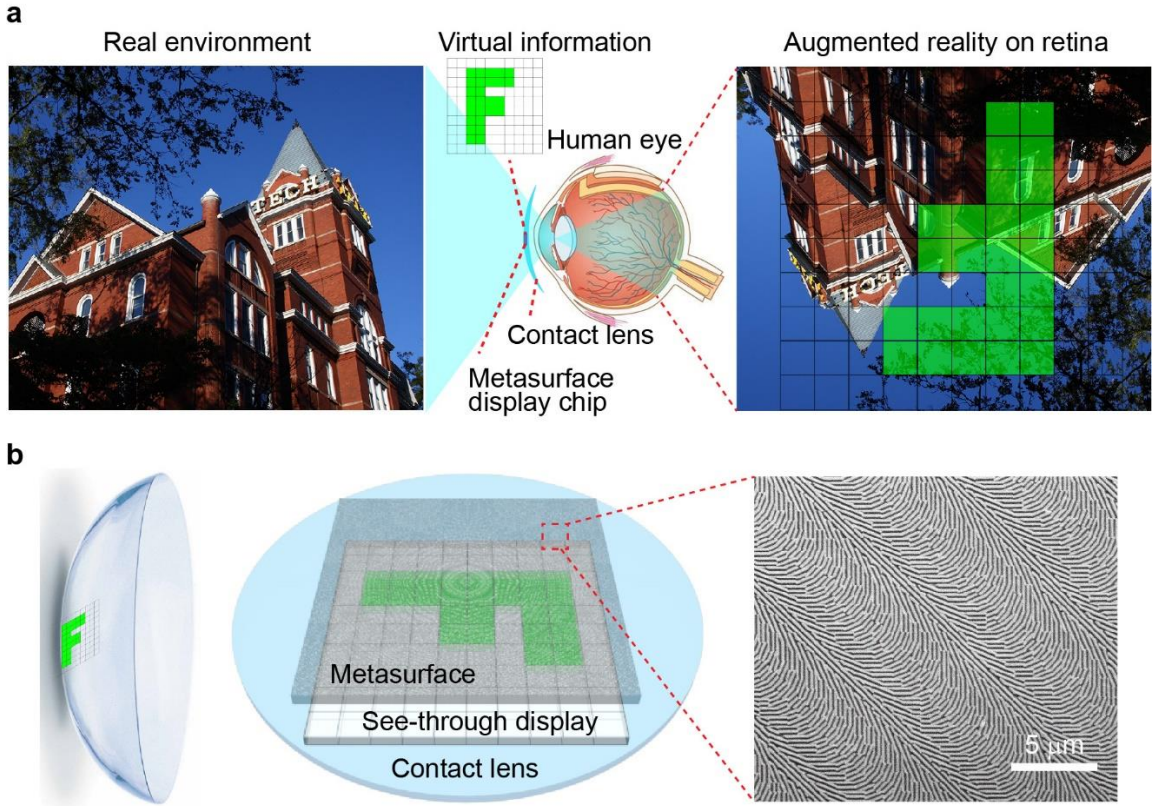

**Figure 1.** Metasurfaces enable AR on contact lenses. **a**, The metasurface on top of a see-through display allows the visualization of virtual information (the letter "F") on a contact lens, which otherwise would not be seen due to the limited accommodation of the human eye. The virtual information coexists with the live view of an unaffected real-world environment, which is represented here as an image of "Tech Tower." **b**, The envisioned AR device contains a see-through display and metasurface situated on a contact lens. The metasurface comprises the predesigned distribution of densely packed silicon nanobeams as Pancharatnam-Berry phase elements shown in the SEM image on the right. The scale bar is 5 μm.

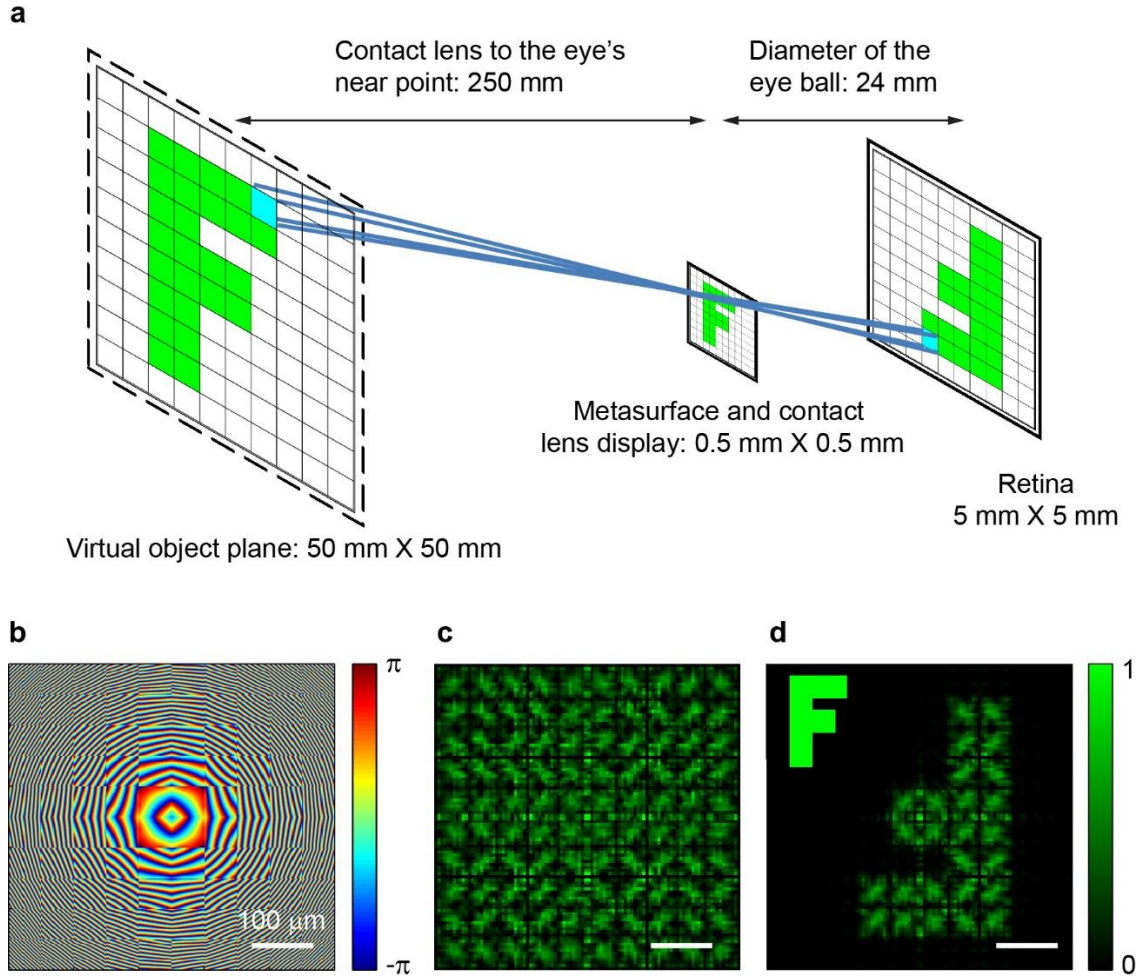

**Figure 2.** Visualize virtual information in a pixel-by-pixel manner. **a**, The visualization of displayed information on the surface of an eyeball. The virtual information that generated on the see-through display transforms into a pixelated pattern at the fovea region of the retina under the aid of the metasurface. With proper accommodation of the eye-lens (not shown), the eye perceives the displayed information appearing at the near point. **b**, Simulated distribution of the phase retardance for holographically casting the received information to an enlarged, pixelated pattern. **c**, All 10 by 10 pixels of the display project onto the retina via the metasurface with the numerically predesigned phase distribution (Fig. 2**b**). **d**, The retina visualizes the pixels of the virtual information, the letter "F" generated on the see-through display (inset), validating the pixel-by-pixel manner of the displaying technique. The scale bars other than specified are 1 mm.

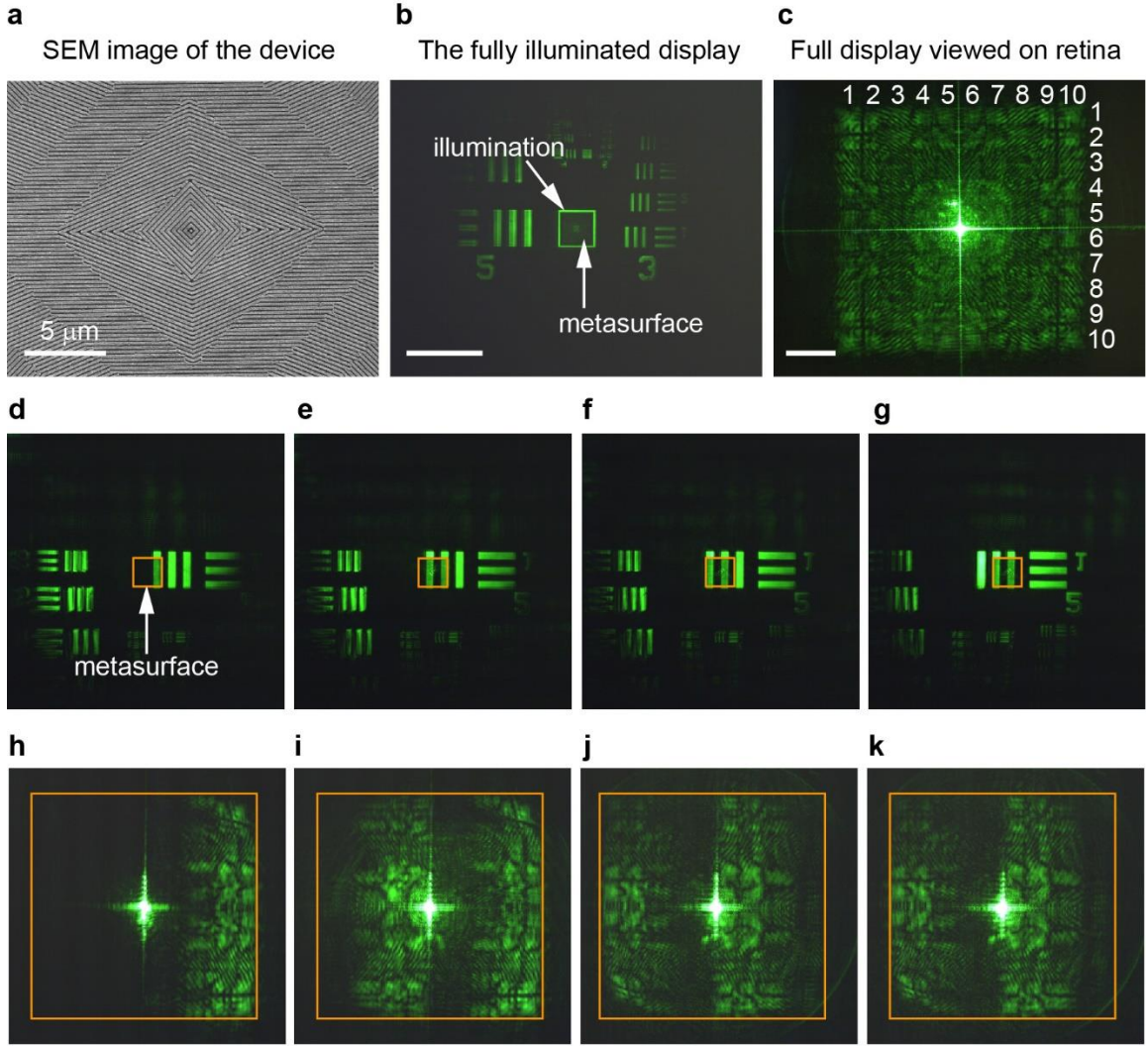

**Figure 3.** Pattern recognition with the metasurface-based AR device. **a**, The SEM image of the metasurface with silicon nano-beams situated at angles (θ) that generate the predesigned phase distribution (φ) in Figure 2b governed by the relationship of φ = 2θ. **b**, The optical image of the metasurface and the virtual information, the green square, on the contact lens shows the full illumination of all pixels. **c**, The retina-camera sees the whole virtual information that comprises 10 by 10 pixels. By illuminating desired pixels, the display generates target patterns (**d** - **g**). Subsequently, mimicked human eye recognizes the patterns (**h** - **k**) under the help of the metasurface to holographically project them onto the retina-camera. In this way, the passive metasurface-based AR device enables the visualization of virtual information on a contact lens by human eyes. The bright cross at the center is the zeroth-order diffraction commonly seen in gradient metasurfaces with geometric phases. The yellow squares are boundaries for guiding eyes.

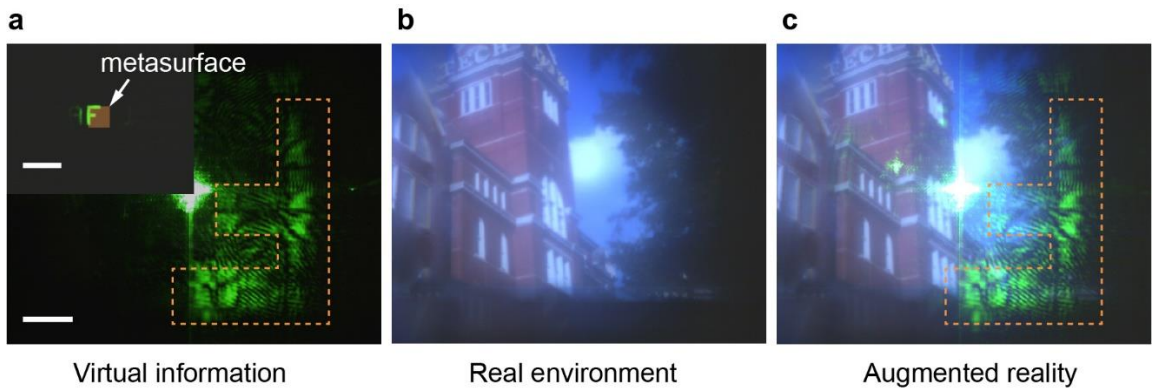

**Figure 4.** Human eye sees the real world with auxiliary virtual information. **a**, The virtual information (inset), letter "F," is an image of the photomask which equivalents to a see-through display on a contact lens. The retina-camera visualizes the pixelated "F," thanks to the ability of the metasurface to holographically projecting the virtual information in a pixel-by-pixel manner. The dotted line is a guide of eye. **b**, The real-world environment, "Tech Tower" in this case, is unaffected in the presence of the metasurface occupying an area of 500 μm by 500 μm, around 1% of a typical pupil area. **c**, The virtual information overlays on the real-world environment, demonstrating AR on contact lenses.

# Metasurfaces for Near-Eye Augmented Reality

*Shoufeng Lan,*[†,‡] *Xueyue Zhang,*[†,‡] *Mohammad Taghinejad,*[†] *Sean Rodrigues,*[†,§] *Kyu-Tae Lee,*[†] *Zhaocheng Liu,*[†] *and Wenshan Cai*[†,§,]\*

[†] School of Electrical and Computer Engineering, Georgia Institute of Technology, Atlanta, Georgia 30332

[§] School of Materials Science and Engineering, Georgia Institute of Technology, Atlanta, Georgia 30332

[‡] Authors contributed equally to this work

\* Correspondence should be addressed to W.C. (wcai@gatech.edu)

**Contents of the Supplementary Information**

S1. Simulation results without predesigned phase distribution

S2. Optical properties of poly-silicon films

S3. Transmission and phase of silicon nanobeams

S4. Schematic of the experimental setup

S5. Displaying complex virtual information

**S1. Simulation results without a predesigned phase distribution**

We have numerically verified (Fig. 2) the needed phase distribution that will be used to holographically cast patterns onto a retina together with a conjugate image at the near point of the eye. We further confirm the proposed imaging method by calculating the intensity distribution on the retina without the predesigned phase distribution. The virtual information, the letter "F" in this case, is the same as that used in Fig. 2. We also keep all other conditions, such as the wavelength of the input light, consistent with that of the main text. Without any phase distribution, the phase is zero for all pixels; the display cannot project the patterns. As a result, the retina only sees a small beam spot (Fig. S1a). We also calculate the intensity distribution with a random phase profile for a metasurface. As expected, the retina sees a randomized intensity distribution (Fig. S1b).

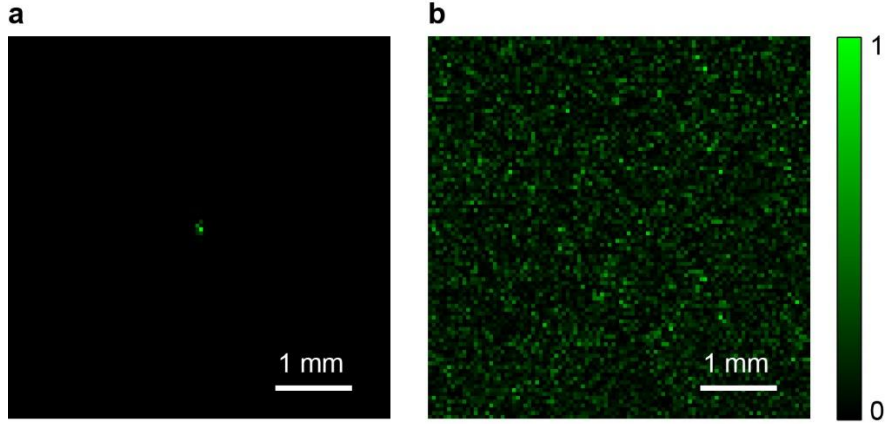

**Figure S1.** Simulated images on the retina without a predesigned phase distribution. **a**, In absence of the metasurface, the retina is not able to see the virtual graphic of "F" on the see-through display, but rather sees a tiny bright dot at the center. **b**, With random phase-distribution of a metasurface, as expected, the retina sees randomized scattering from the virtual graphic on the display.

**S2. Optical properties of poly-silicon films**

We design the geometric parameters, such as the width, thickness, and periodicity, of nanobeams by using experimentally measured optical properties of the poly-silicon for fully converting the input light into the desired pattern while eliminating the zeroth-order diffraction. Ideally, the condition for achieving this is $t_{\text{TE}} = t_{\text{TM}}$ and $\Delta\phi = \pi$ as depicted in the manuscript. The transmission and phase retardance of the nanobeams is governed by the specific geometry of the nanobeams and the optical properties of the poly-silicon film (Fig. S2), which we obtain via ellipsometry measurements. For the measurements, we deposit poly-silicon onto an oxide layer (~400 nm) that is obtained by thermal oxidation of the silicon substrate. Under reflection mode with an oblique incidence, the real and imaginary part of the refractive index for the poly-Si film is extracted via fitting the detected optical signal with theoretical models. In this way, we also obtain the thickness of the film, which is further confirmed by measurements with a profilometer.

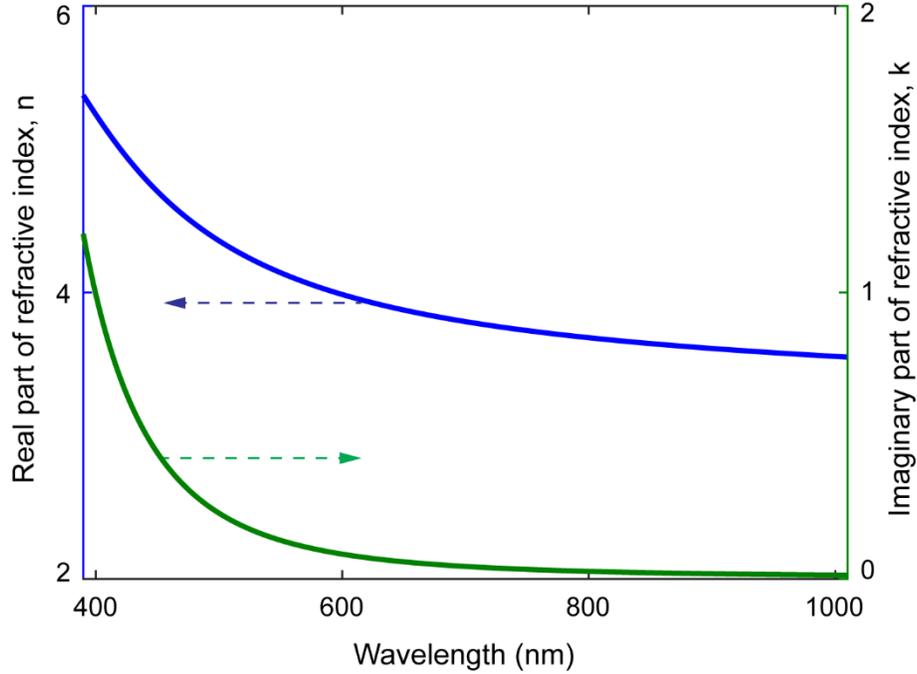

**Figure S2.** Refractive index of the silicon film for the metasurface. The refractive index was measured by a Woollam ellipsometer. The measured data was further applied for the design of the silicon nanobeams.

## S3. Transmission and phase of silicon nanobeams

With the optical properties of the poly-Si film in Fig. S2, we numerically investigate the optical transmission and phase of polarized light through an array of silicon nanobeams (Fig. S3a and S3b), a one-dimensional grating. We simulate the transmission for input light with polarizations both along (TE) and perpendicular to (TM) grating strips, we also extract the phase difference between them ($\Delta\phi$). By varying the geometric parameters, we find the working conditions to be:

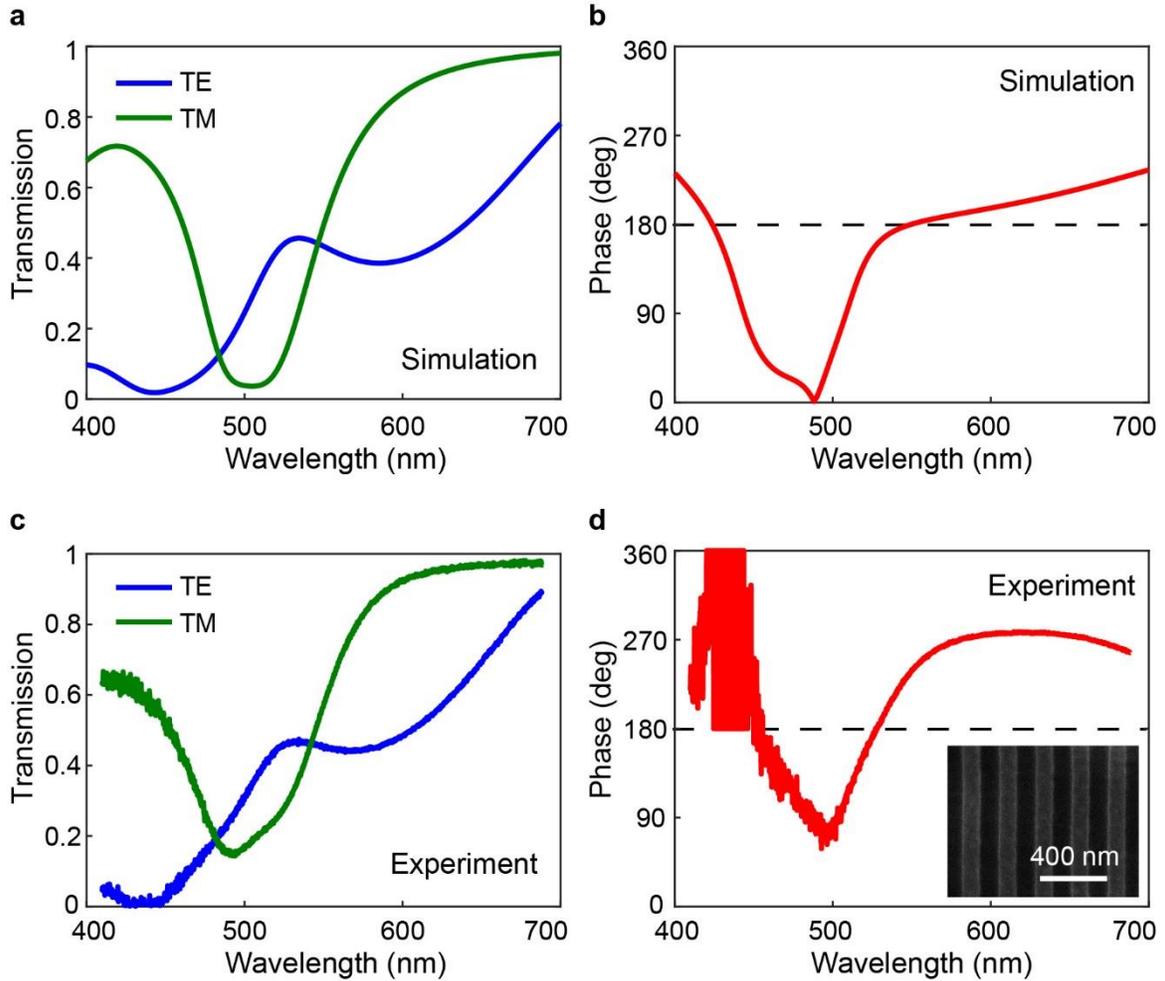

**Figure S3.** Designing the transmission and phase retardance of silicon nanobeams. **a**, The simulated transmission for TE and TM polarizations. The TE (TM) polarization is defined as the electric field component parallel (perpendicular) to the nanobeams. **b**, Simulated phase-retardance between the TE and TM polarizations. The dashed line represents the phase-retardance of 180 degrees. **c**, **d**, Experimentally measured transmission and phase retardance of the metasurface that has the geometric parameters (periodicity is 230 nm, width of the nanobeam is 70 nm) determined by the simulations. The inset is the SEM image of the fabricated silicon nanobeams with a scale bar of 400 nm.

a film thickness of 155nm, a grating width of 70 nm, and a periodicity of 230 nm. With these geometric parameters, the conditions of $t_{\text{TE}} = t_{\text{TM}}$ and $\Delta\phi = \pi$ are fulfilled simultaneously. In principle, the input light fully converts into the output light with a cross polarization, and hence eliminate the zeroth-order diffraction. The experimental results in Fig. S3c and S3d match the simulations relatively well, but not exactly due to fabrication uncertainties. Consequently, the zeroth-order diffraction is not eliminated as shown in the experimental results in Fig. 3 and 4 in the manuscript, which is commonly seen in experimental papers using geometric phases. Moreover, we can easily get rid of the zeroth-order diffraction by placing a polarizer in front of the retina camera because the desired diffraction pattern has a cross-polarization. Nevertheless, the characterizations of transmission and phase for TE and TM polarizations provide a guide for increasing the conversion efficiency with metasurfaces using PB phases.

**S4. Schematic of the experimental setup**

In the experiment, a laser light is circularly polarized by passing through a linear polarizer and a quarter waveplate. We use a photomask to generate an image instead of using a real see-through display to produce virtual information. In other words, the image of the photomask, which locates at the metasurface, is the virtual information for AR in our experiments. Because of the predesigned phase distribution on the metasurface, the image (virtual information) can be captured, or visualized, by the retina-camera with the help of an eye-lens mimicking the real crystalline lens of the human eye. In analogy to the optical system of the human eye, we use a focal length of 17 mm for the eye-lens and a length of 25 mm for the distance between the metasurface and the retina-camera.

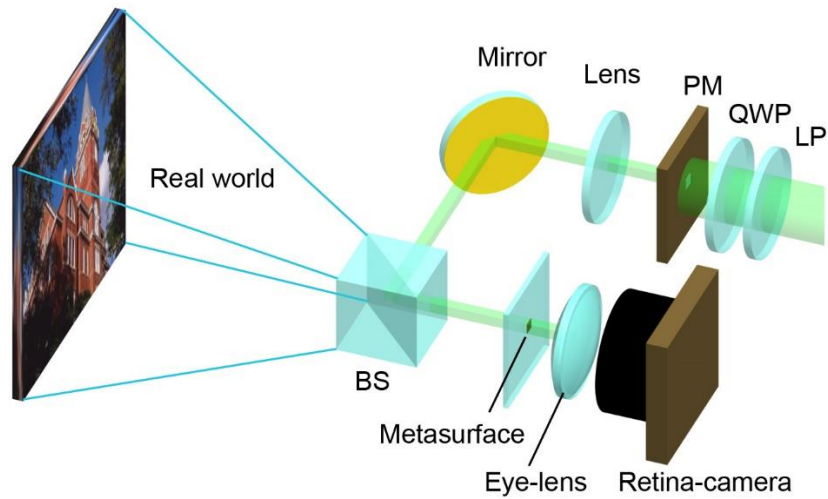

**Figure S4.** Schematic of the experimental setup for AR. LP: Linear polarizer; QWP: quarter waveplate; PM: photomask; BS: beamsplitter.

**S5. Displaying complex virtual information**

With the metasurface-based display containing 10 by 10 pixels in the manuscript, we demonstrate the visualization of letters and some simple patterns such as bars in Fig. 3 on a contact lens. To display more complex patterns, we also numerically design a metasurface containing 100 by 100 pixels in Fig. S5. Using the same strategy as that in the manuscript, we design a phase distribution (Fig. S5a) that empowers the projection of displayed patterns in a pixel-by-pixel manner. Indeed, the retina sees a more complex pattern (Fig. S5b), the American continents, generated from a display (inset) on a contact lens.

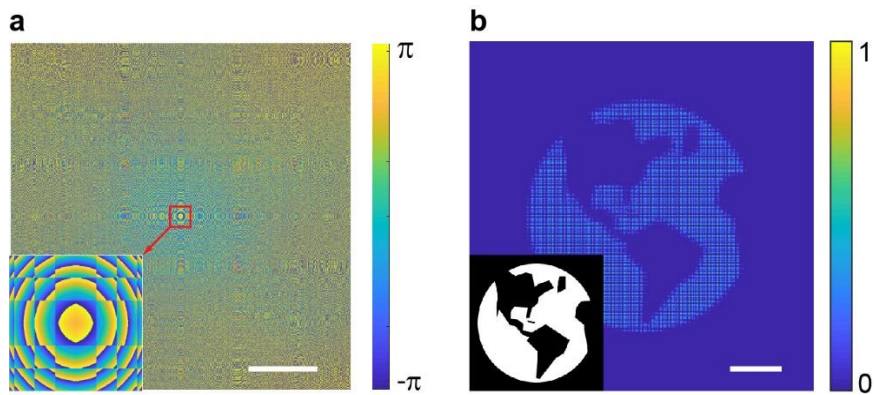

**Figure S5.** A simulated phase distribution with 100 by 100 pixels permits graphics of higher complexity. **a**, The profile of the phase-retardance with 100 by 100 pixels. The inset is a close view of the center region. **b**, The retina sees complex virtual information, the America, with the phase distribution in Fig. S4**a** for the metasurface. The scale bars are 1 mm.